\documentclass[printer]{aa}
\usepackage[varg]{txfonts}
\usepackage{graphicx}
\usepackage{color}
\usepackage{amsbsy,amsmath}
\usepackage{amstext}
\usepackage{natbib}
\usepackage{endnotes}
\newcommand{\be}{\begin{equation}}
\newcommand{\ee}{\end{equation}}
\newcommand{\bea}{\begin{eqnarray}}
\renewcommand{\thefootnote}{\fnsymbol{footnote}}
\newcommand{\eea}{\end{eqnarray}}
\DeclareMathAlphabet\mathbfcal{OMS}{cmsy}{b}{n}

\begin{document}
\title{Tidal evolution of circumbinary systems with arbitrary eccentricities: applications for 
Kepler 
systems}
\titlerunning{Tidal evolution of Kepler circumbinary systems}

\author{F.A. Zoppetti\inst{1,2}, A.M. Leiva\inst{1} and C. Beaug\'e\inst{1,3}}
\authorrunning{F.A. Zoppetti et al.}
\institute{Universidad Nacional de C\'ordoba. Observatorio Astron\'omico de C\'ordoba, Laprida 
854, C\'ordoba X5000GBR, Argentina.\\
\email{fzoppetti@oac.unc.edu.ar}
\and
Consejo Nacional de Investigaciones Cient\'ificas y T\'ecnicas (CONICET), Argentina.\\
\and
CONICET. Instituto de Astronom\'{\i}a Te\'orica y Experimental, Laprida 854, C\'ordoba 
X5000GBR, Argentina.\\}

\abstract{
We present an extended version of the Constant Time Lag analytical approach for the tidal evolution of 
circumbinary planets introduced in our previous work. The model is self-consistent, in the sense 
that all tidal interactions between pairs are computed, regardless of their size. We derive 
analytical expressions for the variational equations governing the spin and orbital evolution, 
which are expressed as high-order elliptical expansions in the semimajor axis ratio but retain 
closed form in terms of the binary and planetary eccentricities. These are found to reproduce 
the results of the numerical simulations with arbitrary eccentricities very 
well, as well as reducing to our previous results in the low-eccentric case.

Our model is then applied to the well-characterised {\it Kepler} circumbinary systems by analysing 
the tidal timescales and unveiling the tidal flow around each different system. In all cases we 
find that the spins reach stationary values much faster than the characteristic timescale 
of the orbital evolution, indicating that all {\it Kepler} circumbinary planets are expected to be 
in a sub-synchronous state. On the other hand, all systems are located in a tidal flow leading 
to outward migration; thus the proximity of the planets to the orbital instability limit may have 
been even greater in the past. Additionally, {\it Kepler} systems may have suffered a significant 
tidally induced eccentricity damping, which may be related to their proximity to the capture 
eccentricity. To help understand the predictions of our model, we also offer a simple 
geometrical interpretation of our results.}

\maketitle
\newpage 

\section{Introduction}
Among the main characteristics of the nine well-characterised circumbinary (CB) systems discovered 
so far by the {\it Kepler} mission, probably the most remarkable one is the proximity of the 
planets to the binaries. This feature implies that most of the circumbinary planets (CBPs) are 
located very close to the stability limit \citep{Holman1999} where, in addition, the in situ 
formation is very unlikely due to perturbations of the binary \citep{Lines2014}.

While it is widely accepted that migration by interaction with a protoplanetary disc took place 
in these systems \citep{Dunhill2013}, it is not yet clear what role tidal forces played in the 
subsequent dynamical evolution, which have the peculiarity of being exerted by two central 
objects of comparable masses and may have had an important effect in such compact systems.

In \cite{Zoppetti2019} we presented a self-consistent weak friction model for tides on a CBP. We 
used Kepler-38 system as a testing case and found several interesting features: for example the planet 
synchronised its spin in a sub-synchronous equilibrium state and, on longer timescales, it migrated 
outward from the binary. We also derived analytical expressions for the variational equations of 
the orbital and spin evolution based on low-order elliptical expansions in semimajor axis ratio 
and eccentricities. This analytical approach was reduced to the two-body case when one of the central 
masses was taken equal to zero and it reproduced the results of our low-eccentric numerical 
simulations very well. However, the low-order expansion in eccentricities of our analytical model did not 
allow its direct application to the eccentric {\it Kepler} systems such as Kepler-34 
\citep{Welsh2012} and Kepler-413 \citep{Kostov2014}.

In this work, we present and discuss an extended version of our self-consistent weak friction 
(Constant Time Lag; CTL) tidal 
model which is valid for any binary and planetary eccentricity. We apply the results to the 
well-characterised nine {\it Kepler} systems in an attempt to reproduce their past tidal 
evolution, as well as unveiling the role played by tidal forces in determining their current 
orbital parameters.

This paper is organized as follows. In Sect. \ref{sec:rev}, we review the construction of the  
variational equations for the spin and orbital evolution in a self-consistent manner. In Sect. 
\ref{sec:anali}, we present the analytical expressions of the variational equations in closed form 
for the eccentricities and compare their predictions with results obtained from numerical 
simulations, as well as with our original low-order expanded model. In Sect. \ref{sec:kep}, we apply our model 
to the {\it Kepler} CB planets by computing the characteristic timescales for their evolution and by 
reproducing the tidal flow around each system. A simple geometrical interpretation of 
the main results of our model is offered in Sect. \ref{sec:inte} to better understand these 
dynamical characteristics. Finally, Sect. \ref{sec:conclu} summarises and discusses our main 
results.

\section{Review of the model}
\label{sec:rev}

We return to the problem of a binary system with stellar masses $m_0$ and $m_1$, and a CBP with 
mass $m_2$. The binary and the planet share the same orbital plane and their spins are 
perpendicular to it. We further assume that each body is an extended mass with physical 
radius $\mathcal{R}_i$ and deformable due to the tidal effects of its companions. 

We again assume that the internal dissipation due to tides in each of the three bodies is governed by the CTL
and is, therefore, subject to the description suffested by \cite{Mignard1979}. In our previous work 
\citep{Zoppetti2019} we demonstrated that, in absence of any mean-motion commensurability between 
$m_1$ and $m_2$, the cross tides can be neglected and the only tidal forces that need to be
taken into account on each body $m_i$ are 
those caused by the deformation that the mass $m_i$ itself induces on its companions. For this 
reason, in a multi-body tidal model like the one we are studying, the tides on each body can be 
simply computed by summing the forces and torques between deforming-deformed pairs as given in 
Equations (5) and (6) of \cite{Mignard1979}.

In an inertial frame with arbitrary origin, where the position of each body $m_i$ is denoted as 
${\bf R}_i$, the orbital evolution equations may be expressed as the sum of the point-mass and 
tidal interactions, which is
\bea
\begin{aligned}
m_i {\bf \ddot{R}}_i  &= \;\;\, 
\sum_{j=0 ,\ j\neq i}^2 \bigg( \frac{\mathcal{G}m_i \, m_j}{|{{\bf{\Delta}}_{ji}}|^3} {{\bf 
\Delta}_{ji}} + ({{\bf{F}}_{ij}}-{{\bf{F}}_{ji}}) \bigg),   \ \ \ \ \ \ \ i= 0,1,2
\label{eq:orb}
\end{aligned}
\eea
where $\mathcal{G}$ is the gravitational constant and ${\bf \Delta}_{ij} \equiv {\bf R}_i - {\bf 
R}_j$. We denoted by ${{\bf{F}}_{ij}}$ the tidal force acting on $m_i$ due to the deformation that 
it induces on $m_j$ and, following \cite{Ferraz-Mello2008}, we also took into account the reacting 
forces when computing the tides on each mass $m_i$. According to \cite{Mignard1979}, these can be 
expressed by 
\bea
{{\bf{F}}_{ij}} = -\frac{\mathcal{K}_{ij}} {{|{{\bf \Delta}_{ij}}|}^{10}}
\bigg[ 2({{\bf \Delta}_{ij}} \cdot {{\bf \dot{\Delta}}_{ij}}) {{\bf \Delta}_{ij}} 
+ {{\bf \Delta}^2_{ij}} ( {{\bf \Delta}_{ij}} \times {{\bf{\Omega}}_j} + {{\bf \dot{\Delta}}_{ij}} 
)  
\bigg] 
\label{eq:mig}
\eea
where ${\bf \Omega}_j$ is the spin vector of $m_j$ (assumed parallel to the orbital angular 
momenta) and the magnitude factor of the tidal force $\mathcal{K}_{ij}$ is given by
\be
\mathcal{K}_{ij} = 3 \mathcal{G}m_i^2 \mathcal{R}_j^5 k_{2j}\Delta t_j,
\label{eq:k}
\ee
where $k_{2j}$ and $\Delta t_j$ are the second degree Love number and the time-lag of the body 
$m_j$. Note that we have neglected the contribution of the static tides since, 
in the absence of any mean-motion resonance, it does not introduce any angular momentum exchange and
therefore yields no secular change in the spins nor in the orbital evolution.

While expression (\ref{eq:orb}) governs the orbit, the rotational dynamics can be deduced from the 
conservation of the total angular momentum. Assuming that the variation in the spin angular 
momenta of the body $m_i$ is only due to the contributions associated to its own deformation, we 
can uncouple the spin evolution equations of each body to obtain 
\be
\frac{d{\bf \Omega_i}}{dt} = \frac{1}{{\mathcal C}_i} \sum_{j \ne i} 
\frac{{\cal K}_{ji}}{|{{\bf \Delta}_{ji}}|^6} \bigg[ \frac{{{\bf \Delta}_{ji}} \times 
{{\bf {\dot \Delta}}_{ji}}}{|{{\bf \Delta}_{ji}}|^2} - {{\bf \Omega}_i} \bigg],   \ \ \ \ \ \ \ i= 0,1,2 
\label{eq:rot}
\ee
where $\mathcal{C}_i$ is the principal moment of inertia of $m_i$.

\section{Analytical secular model for arbitrary eccentricities }
\label{sec:anali}

To construct the analytical model of the orbital evolution Equations (\ref{eq:orb}) and the 
rotational evolution Equations (\ref{eq:rot}), we first adopt a Jacobi reference frame for the 
position and velocities of the bodies. In this system, the Jacobi position vectors of the masses 
${\boldsymbol \rho}_i$ can be expressed in terms of their inertial counterparts ${\bf R}_i$ as
\bea
\begin{aligned}
{\boldsymbol \rho_{ 0}} &= \frac{1}{\sigma_2} (m_0 \, {\bf R}_0 + m_1 \, {\bf R}_1 + m_2 \, 
{\bf R}_2) \\
{\boldsymbol \rho_{ 1}} &= {\bf R}_1 - {\bf R}_0  \\
{\boldsymbol \rho_{ 2}} &= {\bf R}_2  - \frac{1}{\sigma_1} (m_0 \, {\bf R}_0 + m_1 \, {\bf 
R}_1) ,\\
\end{aligned}
\label{eq:jac}
\eea
where $\sigma_i=\sum_{k=0}^i m_k$. The transformation equations for the velocities are analogous.

To construct an analytical secular model valid for any binary and planetary eccentricity, 
we note that both (\ref{eq:orb}) and (\ref{eq:mig}) depend on powers of the inverse of 
the relative distances, i.e. ${|{{\bf \Delta}_{ij}}|}^{(-n)} \equiv {|{{\bf R}_{j} - {\bf 
R}_{i}}|}^{(-n)}$, with $n$ a positive integer. Assuming momentarily that $t = R_i/R_j < 1$, we 
can write
\be
g^{(n)}(t,x) \equiv \left( {\frac{R_j}{|{{\bf \Delta}_{ij}}|}} \right)^{n} = \bigl( 1 + t^2 - 2tx
\bigr)^{-n/2}
\ee
where $x$ is the cosine of the angle between both position vectors. The expansion of $g^{(1)}(t,x)$ in power 
series of $t$ is well known
\be
g^{(1)}(t,x) = \sum_{k=0}^{\infty} t^k \, P_k(x)
\ee
with $P_k(x)$ the $k$-degree Legendre polynomial. In a similar manner, for $n>1$ we can introduce 
`generalized' polinomials $Z^{(n)}_k(x)$ such that 
\be
g^{(n)}(t,x) = \sum_{k=0}^{\infty} t^k \, Z^{(n)}_k(x)
\ee
such that $Z^{(0)}_k(x) = 1$ and $Z^{(1)}_k(x) = P_k(x)$ for all values of $k \ge 0$. These new 
polynomials can be obtained for the original Legendre functions through recurrence relations. 
Writing $g^{(n)}(t,x) = g^{(n-1)}(t,x) \cdot g^{(1)}(t,x)$, replacing each function by its series 
expansion and equating equal powers of $t$, we find
\be
Z^{(n)}_k(x) = \sum_{j=0}^k P_{k-j}(x) \,\, Z^{(n-1)}_j(x) . 
\ee

\begin{figure*}[ht!]
\centering
\includegraphics[width=1.99\columnwidth,clip]{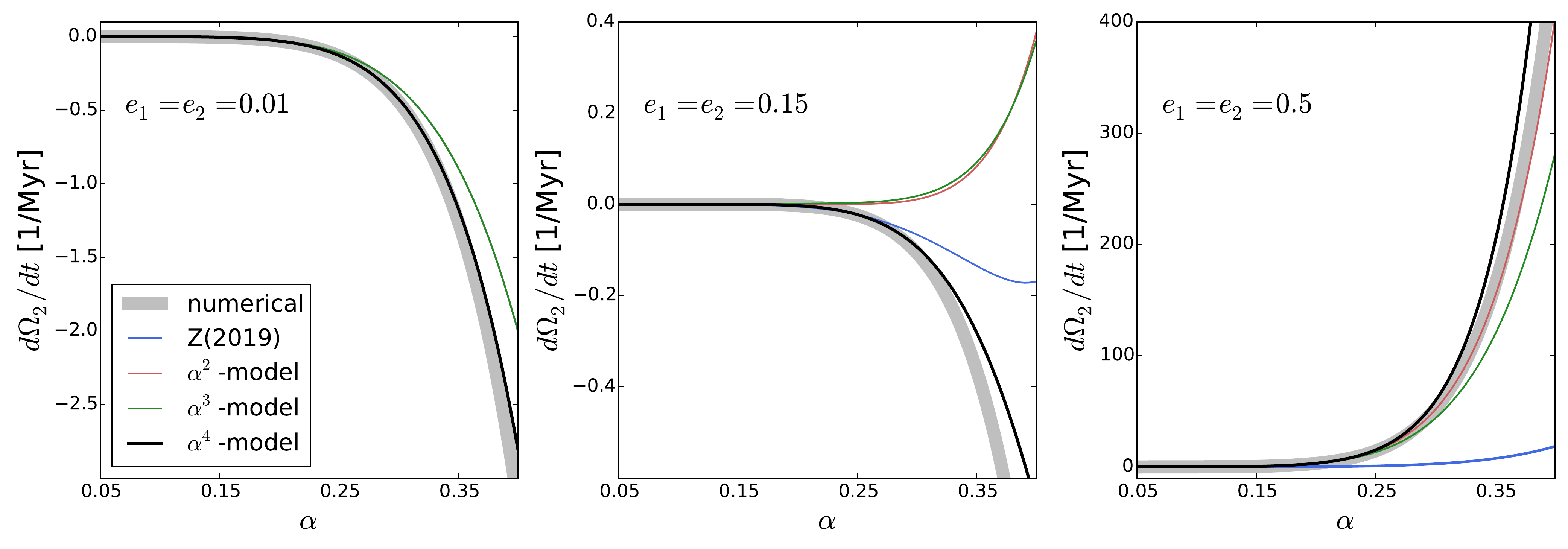}
\caption{Time derivative of the planetary spin rate $\Omega_2$ in our generic binary system as a 
function of $\alpha=a_1/a_2$. Each panel assumes different eccentricities for both orbits. Broad 
gray curves show results from semi-analytical calculations and predictions from analytical 
models are depicted in thin lines. Values obtained with our original \citep{Zoppetti2019} are 
presented in blue, while the colors correspond to our current model with different orders 
in $\alpha$. See inlaid box in left-hand plot for details.}
\label{fig:do2dt}
\end{figure*}

Introducing these expressions into the equations of motion, transforming the position and 
velocity vectors to Jacobi variables, and averaging over the mean longitudes $\lambda_1$ and 
$\lambda_2$, we find closed analytical expressions in terms of the eccentricities $e_1$ and 
$e_2$. The procedure is analogous to that followed by \cite{Mignard1980} in the two-body case, with 
two important exceptions. First, we consider expansions in $\alpha=a_1/a_2$ up to fourth order 
(i.e. $k=4$); second, we do not average over $\varpi_1$ and $\varpi_2$. While this would greatly 
simplify the resulting equations, in \cite{Zoppetti2019} we observed that in some simulations, 
typically associated to low planetary eccentricity, the planet and the secondary star entered in 
the aligned secular mode $\Delta\varpi=\varpi_2-\varpi_1 \sim 0$. Thus, the perturbation terms 
dependent on $\Delta \varpi$ may have a net contribution in the long-term dynamics, at least for 
some initial conditions.

The complete procedure was carried out with the aid of an algebraic manipulator. Some intermediate 
expressions had to be expanded up to fifth order in $t = R_i/R_j$ to avoid loss of precision in 
the final results. The process was complex, especially for $k > 2$ where we found little benefit 
explicitly working with the $Z^{(n)}_k(x)$ functions. However, we keep their definition and 
recurrence relations since they could be useful for quadrupole-level theories.

\subsection{Planetary spin evolution}
\label{anali_spin}
The process discussed previously was applied to both the orbital and spin variational equations. 
In the latter case, up to fourth order in semimajor axes ratio, we obtain
\be
\left<\frac{d\Omega_2}{dt}\right> = \frac{3\mathcal{G}\mathcal{R}_2^5k_{2,2}\Delta t_2}{2 \mathcal{C}_2 
a_2^6 } \sum_{i=0}^4 K^{(s)}_i \bigg( A^{(s)}_{i} n_1 + B^{(s)}_{i} n_2 + C^{(s)}_{i} \Omega_2 
\bigg) \, \alpha^i ,
\label{eq:dodt}
\ee
where $n_1$ is the mean motion of the binary, $n_2$ that of the planet, and
\be
\label{eq:ks}
K^{(s)}_i = \gamma_0^i m_0^2 + \gamma_1^i m_1^2 \ \ \ , \ \ \
\gamma_0 = \frac{m_1}{\sigma_1} \ \ \ , \ \ \ \gamma_1 = -\frac{m_0}{\sigma_1} .
\ee
The coefficients $A_i^{(s)}$, $B_i^{(s)}$ and $C_i^{(s)}$, functions of the eccentricities and 
difference of longitudes of pericenter, are listed in Appendix \ref{ap:crot}. 

To test the accuracy of our analytical model, we consider a generic circumbinary system with 
primary mass $m_0=1 M_\odot$ and physical radius $\mathcal{R}_0 = 1 \mathcal{R}_\odot$, plus a 
secondary star with mass $m_1 = m_0/3$ and a physical radius $\mathcal{R}_1 = 
(m_1/M_\odot)^{0.9} \mathcal{R}_\odot$ \citep{Demircan1991}. The mass parameter of the 
binary is thus $\bar{\mu}=m_1/(m_0+m_1)=0.25$ and the semimajor axis was set to $a_1 = 0.15$ AU.
For the planet, we chose $m_2 = 10 M_\oplus$ and $\mathcal{R}_2=4 \mathcal{R}_\oplus$. We assumed 
a planetary principal moment of inertia equal to $\mathcal{C}_2 = 0.25 m_2 \mathcal{R}_2^2$ and 
a modified tidal dissipation factor of $Q_2' = 3/(2 n_2 k_{2,2}\Delta t_2) = 10$.

Figure \ref{fig:do2dt} shows the time derivative of $\Omega_2$ as a function of $\alpha=a_1/a_2$
for three different sets of eccentricities. Different curves are used to show results using 
different models. Values obtained by numerically averaging Equation (\ref{eq:dodt}) are depicted 
in broad grey curves, the analytical model from \cite{Zoppetti2019} is shown in blue curves, 
and different variations of our current model are presented in other colors. We observe that our 
new $\alpha^4$ model fits the numerical values very well up to $\alpha \sim 0.35$. The accuracy of 
our closed-form expressions in the eccentricities is evident when we analyze cases in which 
at least one of the orbits is moderately or highly eccentric. These situations were beyond the 
scope of our previous model. The low-eccentricity regime is well reproduced in all 
models, although again we obtain more precision with the closed fourth-order approach.

Curiously, due to the complexity of the power series, results obtained with $\alpha^3$ are not 
always better than those obtained with a quadrupole approximation. Moreover, in some cases 
(e.g. the central panel) both of these models yield predictions which are less accurate compared to 
those obtained from \cite{Zoppetti2019}. These results show the importance of using a high 
order expansion in semimajor axis ratio, even for systems that are not particularly 
compact. 

The value of the stationary spin rate $\left< \Omega_2 \right>^{(\rm st)}$ predicted by our 
model can be calculated equating Equation (\ref{eq:dodt}) to zero, which yields: 
\begin{eqnarray}
\left< \Omega_2 \right>^{(\rm st)} = - \frac{\sum_{i=0}^4 K_{i}^{(s)} D_i^{(s)} 
\alpha^i}{\sum_{i=0}^4 K_{i}^{(s)} C_i^{(s)} \alpha^i},
\end{eqnarray}
where we have denoted $D_i^{(s)} = A_i^{(s)} n_1 + B_i^{(s)} n_2$. Expanding this expression in 
power series of $\alpha$ up to fourth order gives
\be
\left< \Omega_2 \right>^{(\rm st)} = \sum_{i=0}^4 \left< \Omega_2 \right>^{(\rm st)}_i \alpha^i, 
\label{eq:omeq_ana}
\ee
where the coefficients $\left< \Omega_2 \right>^{(\rm st)}_i$ are again listed in Appendix 
\ref{ap:crot}.

\begin{figure}
\centering
\includegraphics[width=1.\columnwidth,clip]{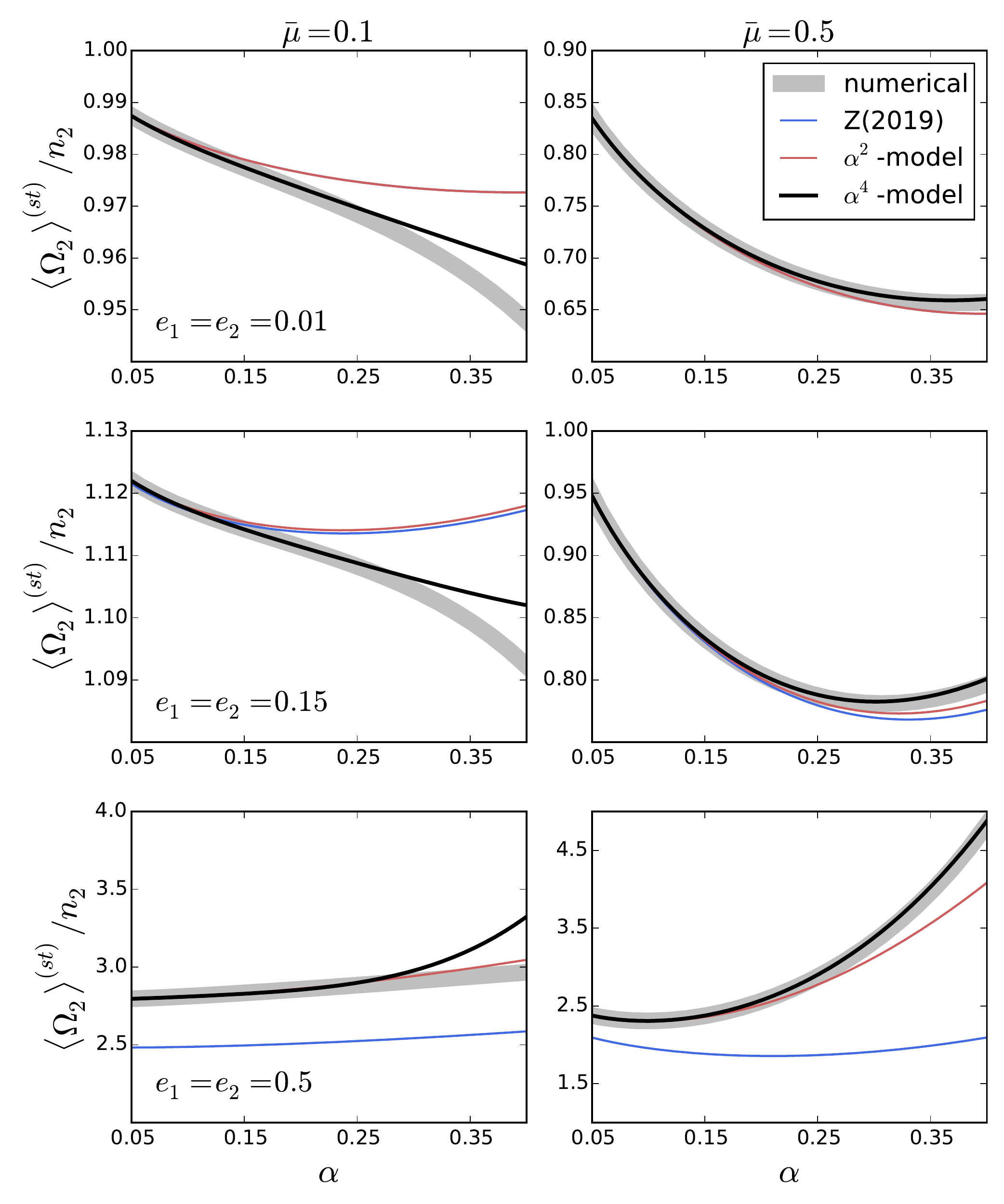}
\caption{Stationary spin of a CB planet as a function of $\alpha$. Different rows consider 
different eccentricity of the orbits (assumed equals). Different columns represent different 
reduced mass: $\bar{\mu}=0.1$ (left panels) and $\bar{\mu}=0.5$ (right panels). The type of the 
curves represent the same that in Figure \ref{fig:do2dt}.}
\label{fig:ome_eq}
\end{figure}

\begin{figure}
\centering
\includegraphics[width=1.\columnwidth,clip]{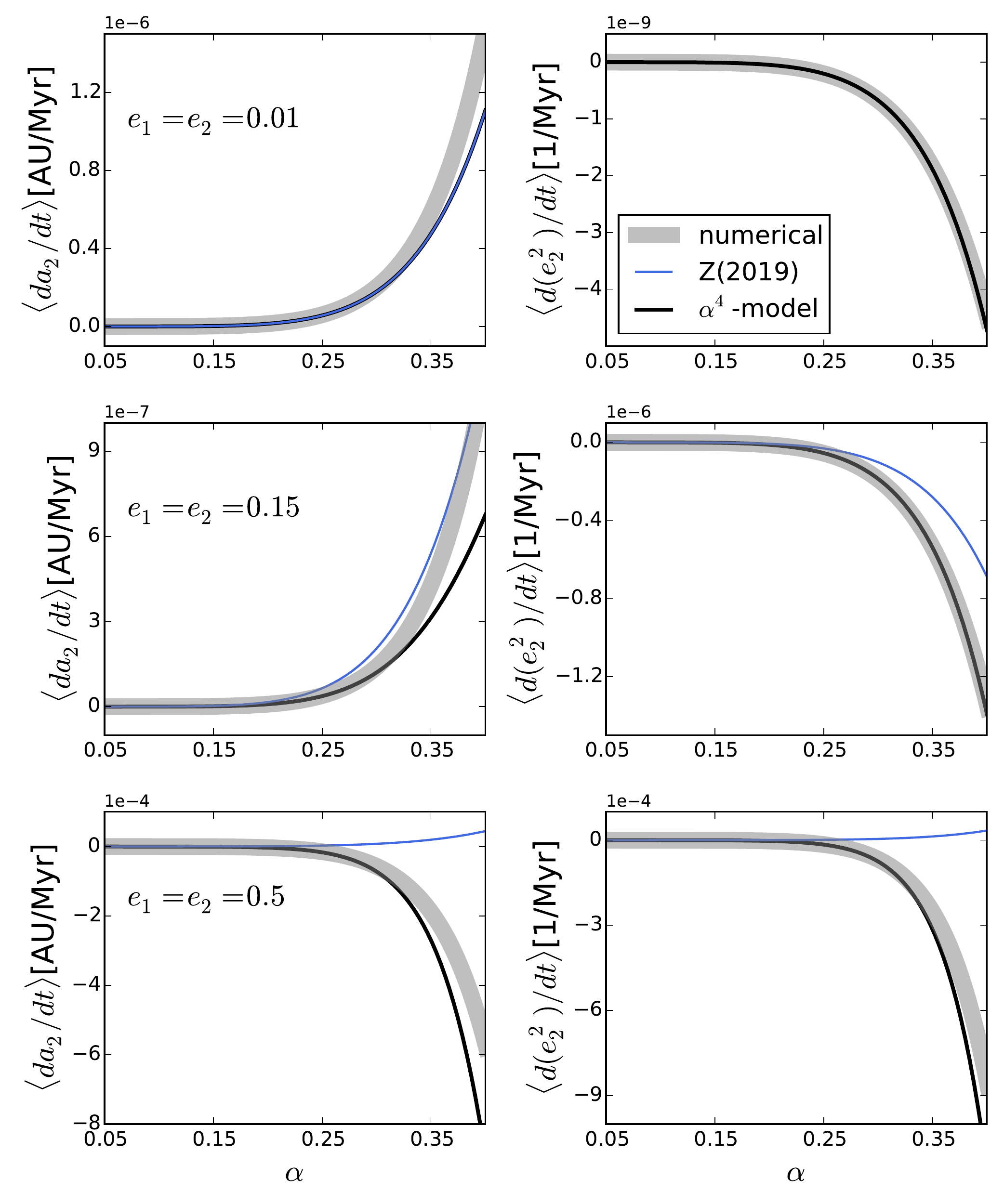}
\caption{Time derivative of the planetary semimajor axis (left panels) and eccentricity (right 
panels), as a function of $\alpha$, for a synchronised CB planet around our generic binary 
system. As before, results obtained with different models are shown in different colored 
curves: numerical in broad grey, \cite{Zoppetti2019} in thin blue, and predictions from our 
current $\alpha^4$ model in black.}
\label{fig:dade}
\end{figure}

Figure \ref{fig:ome_eq} shows the stationary spins (divided by $n_2$), as function of $\alpha$, 
calculated with the same analytical models considered in Figure \ref{fig:do2dt}. To avoid too 
much clutter, we did not plot results corresponding to the $\alpha^3$ model. We also considered two 
different values of the binary mass parameter: results for $\bar{\mu}=0.1$ are shown in the left 
panels while those obtained for $\bar{\mu}=0.5$ are graphed in the right panels.

Comparisons with numerical estimations (broad gray curves) show that our current $\alpha^4$ model 
predict reliable stationary solution for any binary and planetary eccentricity up to $\alpha \sim 
0.35$. Larger mass ratios appear to require less sophisticated dynamical models, while at least a 
$\alpha^4$ model is necessary for low-mass secondary components. We also note that the mass of the 
secondary star (represented by $\bar{\mu}$) competes against the planetary eccentricity in 
establishing the stationarity solution: higher secondary masses tend to sub-synchronous 
solutions while, as in the two-body case, high planetary eccentricity leads to super-synchronous 
solutions. This fact was previously observed in \cite{Zoppetti2019} for low-eccentric systems.

\subsection{Secular evolution for the planetary semimajor axis and eccentricity}
\label{sec:aye}

The variational equation for the planetary semimajor axis in the Jacobi frame is given by
\citep{Beutler2005}
\be
\label{eq:dadt}
\frac{da_2}{dt} = \frac{2a_2^2}{\mathcal{G}\sigma_2} (\dot {{\boldsymbol \rho}_2} \cdot \delta 
{\bf f}_2 )
\ee
where $\delta {\bf f}_2$ is the total tidal force (per unit mass) affecting the two-body motion of 
the planet around the barycenter of the binary system. It is explicitly given by
\be
\delta {\bf f}_2 = \frac{1}{\beta_2}\bigg[ ({\bf F}_{2,0}-{\bf F}_{0,2})+({\bf F}_{2,1}-{\bf 
F}_{1,2})   \bigg],
\ee
where $\beta_2 = m_2 \sigma_1/\sigma_2$ is the Jacobian reduced mass of the planet. We then repeat 
the procedure employed in Section 3.1, now in order to obtain averaged expressions for the 
orbital evolution which are in closed form with respect to the eccentricities. Up the fourth order 
in $\alpha$ and averaging over the mean longitudes, we obtain
\be
\label{eq:dadt_ana}
\left<\frac{da_2}{dt}\right> = \frac{ n_2}{\mathcal{G}m_2\sigma_1 a_2^4} \sum_{i=0}^4 
K_i^{(a)}\bigg(  A^{(a)}_{i} n_1 + B^{(a)}_{i} n_2  + C^{(a)}_{i} \bar\Omega_i^* \bigg) \; 
\alpha^i,
\ee
where 
\be
\label{eq:Ka}
K_i^{(a)} = \mathcal{K}_0^{(+)}\gamma_0^i+\mathcal{K}_1^{(+)}\gamma_1^i \ \ \ , \ \ \
\mathcal{K}_j^{(+)} = \mathcal{K}_{2,j} + \mathcal{K}_{j,2},
\ee
and the coefficients $A_i^{(a)}$, $B_i^{(a)}$, and $C_i^{(a)}$, are listed in Appendix \ref{ap:csem}.
The variational equation for the semimajor axis depends on the spins through a new complex 
parameter given by
\be
\label{eq:O*}
\bar\Omega_i^* = 
\frac{(\mathcal{K}_{2,0}\Omega_0+\mathcal{K}_{0,2}\Omega_2)\gamma_0^i+(\mathcal{K}_{2,1}
\Omega_1+\mathcal{K}_{1,2}\Omega_2)\gamma_1^i}{\mathcal{K}_0^{(+)}\gamma_0^i+\mathcal{K}_1^{(+)}
\gamma_1^i} .
\ee

Finally, the time evolution of the planetary eccentricity may be found from the orbital angular 
momentum 
\be
L_2= \sqrt{{\bf{L}}_2 \cdot {\bf{L}}_2} =\beta_2\sqrt{\mathcal{G}\sigma_2a_2(1-e_2^2)}.
\ee 
Differentiating over time and extracting the eccentricity term, we may write
\be
\frac{d}{dt} (e^2_2) = \frac{1}{a_2} \bigg[(1-e^2_2)\frac{da_2}{dt} - \frac{2 
L_2}{\mathcal{G}\sigma_2\beta_2} \bigg( ({\boldsymbol{\rho}}_2 \times \delta{\bf f}_2) \cdot 
\hat{\bf z}_2\bigg) \bigg],
\label{eq:dedt}
\ee
where $\hat{{\bf z}}_2$ is the unit vector of the orbital angular momentum. Proceeding in the same 
manner that for Equation (\ref{eq:dadt}), the rate of change of the planetary eccentricity can be 
approximated as:
\be
\label{eq:dedt_ana}
\left<\frac{de_2^2}{dt}\right> = \frac{ n_2}{\mathcal{G}m_2\sigma_1 a_2^5} \sum_{i=0}^4 
K_i^{(a)}\bigg(  A^{(e)}_{i} n_1 + B^{(e)}_{i} n_2  + C^{(e)}_{i} \bar\Omega_i^* \bigg) \; 
\alpha^i. 
\ee
We note that both the tidal magnitude coefficient $K_i^{(a)}$ and the spin functions $\bar\Omega_i^*$ 
are the same as deduced for the semimajor axis. The coefficients $A_i^{(e)}$, $B_i^{(e)}$, 
and $C_i^{(e)}$, are listed in Appendix \ref{ap:csem}.

Figure \ref{fig:dade} shows the derivative of the semimajor axis (left panels) and eccentricity 
(right panels), as a function of $\alpha$, of a CBP orbiting our generic binary system. The 
spin rates of the stars and the planet were set in their stationary values: $\Omega_0=\Omega_1 
\simeq n_1 (1+6e_1^2)$ \citep{Hut1980} and $\Omega_2 = \left< \Omega_2 \right>^{(\rm st)}$ 
(Equation (\ref{eq:omeq_ana})). We observe that our closed fourth-order model is a very precise 
approximation of the numerical values up to $\alpha \sim 0.35$. The need of an analytical model 
closed in eccentricities is again particularly evident in the high-eccentricity regime (bottom 
plots), where our previous model fails not only to estimate the magnitude but may also predict 
erroneous signs for the time derivative.

Regarding the dynamical implications, we observe that tides always seem to decrease the 
planetary eccentricity, with magnitude proportional to $e_2$ and also to the proximity to the 
binary. The effect of tides on the planetary semimajor axis is more diverse: for low-to-moderate 
planetary eccentricities, the tidal forces leads to an outward migration, while inward migration is 
expected for the high-eccentricity case. These results confirm our previous findings 
\citep{Zoppetti2019} and a simple geometrical explanation is offered in Section 4.2.

\section{Application to Kepler systems}
\label{sec:kep}

Having developed a model valid for arbitrary binary and planetary eccentricities, we can analyze 
the tidal evolution of {\it Kepler} CB systems.

\begin{table*}
\centering
\caption{Physical and orbital parameters of {\it Kepler} circumbinary systems. Blank spaces
indicate unknown values.}
\label{tab:kep}
\begin{spacing}{1.5}
\resizebox{0.95\textwidth}{!}{%
\begin{tabular}{c|c|c|c|c|c|c|c|c|c|c|c}
\hline \hline
               &   $m_0 (M_\odot)$   &   $m_1 (M_\odot)$   &   $m_2 (M_\oplus)$   &   $\mathcal{R}_0 (\mathcal{R}_\odot)$   &   $\mathcal{R}_1 (\mathcal{R}_\odot)$   &   $\mathcal{R}_2 (\mathcal{R}_\oplus)$   &   $a_1 (AU)$   &   $a_2 (AU)$   &   $e_1$   &   $e_2$ & Ref. \\
\hline
Kepler-16      & 0.6897 & 0.20255 & 105.86 & 0.6489 & 0.22623 & 8.449 & 0.22431 & 0.7048 & 0.15944 & 0.0069 & \cite{Doyle2011}  \\

Kepler-34      & 1.0479 & 1.0208 & 69.94 & 1.1618 & 1.0927 & 8.564 & 0.22882 & 1.0896 & 0.52087 & 0.182 & \cite{Welsh2012} \\

Kepler-35      & 0.8877 & 0.8094 & 40.37 & 1.0284 & 0.7861 & 8.160 & 0.17617 & 0.60347 & 0.1421 & 0.042 & \cite{Welsh2012} \\

Kepler-38      & 0.949 & 0.249 & <122 & 1.757 & 0.2724 & 4.35 & 0.1469 & 0.4644 & 0.1032 & <0.032 & \cite{Orosz2012} \\

Kepler-47b      & 0.957 & 0.342 & <26 & 0.936 & 0.338 & 3.05 & 0.08145 & 0.2877 & 0.0288 & 0.0210 & \cite{Orosz2012b} \\

Kepler-47d      &  &  & 7-43 &  & & 7.04 & & 0.6992 &  & 0.024 & \cite{Orosz2019} \\

Kepler-47c      &  &  & 2-5 &  &  & 4.65 &  & 0.9638 &  & 0.044 & \cite{Orosz2012b}  \\

Kepler-64      & 1.528 & 0.408 & <169 & 1.734 & 0.378 & 6.18 & 0.1744 & 0.652 & 0.2117 & 0.0702 & \cite{Schwamb2013} \\

Kepler-413     & 0.820 & 0.5423 & 67.0 & 0.7761 & 0.484 & 4.347 & 0.10148 & 0.3553 & 0.0365 & 0.1181 & \cite{Kostov2014}\\

Kepler-453     & 0.944 & 0.1951 & <16 & 0.833 & 0.2150 & 6.204 & 0.18539 & 0.7903 & 0.0524 & 0.0359 & \cite{Welsh2015}\\

Kepler-1647    & 1.2207 & 0.9678 & 483.0 & 1.7903 & 0.9663 & 11.8739 & 0.1276 & 2.7205 & 0.1602 & 0.0581 & \cite{Kostov2016} \\

\hline
\end{tabular}}
\end{spacing}
\end{table*}

\begin{table*}
\centering
\caption{Tidal stationary spin rates and characteristic timescales ({\bf{assuming $Q'_2=1000$}}) for 
{\it Kepler} circumbinary systems.}
\label{tab:kep_tid}
\begin{spacing}{1.5}
\resizebox{0.75\textwidth}{!}{%
\begin{tabular}{c|c||c|c||c|c||c|c}
\hline \hline
& $\alpha$ & $\tau_s$ (yr) & $\left<\Omega_2\right>^{(\rm st)}/n_2$ &  $\left< da_2/dt \right>$ 
(AU/yr) & $\tau_a$ (yr) & $\left< de_2/dt \right>$ (1/yr) & $\tau_e$ (yr) \\
\hline
Kepler-16   & 0.318 & $2.2\times10^{7}$ & 0.849 & $4.1\times10^{-16}$ & $1.7\times10^{15}$ & 
$-1.9\times10^{-16}$ & $3.7\times10^{13}$  \\
Kepler-34   & 0.210 & $2.6\times10^{7}$ & 0.947 & $2.0\times10^{-16}$ & $5.6\times10^{15}$ & 
$-4.5\times10^{-16}$ & $4.1\times10^{14}$  \\
Kepler-35   & 0.292 & $2.0\times10^{6}$ & 0.683 & $9.7\times10^{-15}$ & $6.2\times10^{13}$ & 
$-3.2\times10^{-15}$ & $1.3\times10^{13}$  \\
Kepler-38   & 0.316 & $1.4\times10^{6}$ & 0.875 & $2.3\times10^{-15}$ & $2.0\times10^{14}$ & 
$-3.3\times10^{-15}$ & $9.6\times10^{12}$  \\
Kepler-47b  & 0.283 & $4.5\times10^{5}$ & 0.818 & $8.0\times10^{-15}$ & $3.6\times10^{13}$ & 
$-6.6\times10^{-15}$ & $3.2\times10^{12}$  \\
Kepler-47c  & 0.116 & $5.4\times10^{6}$ & 0.880 & $9.7\times10^{-16}$ & $7.2\times10^{14}$ & 
$-5.8\times10^{-16}$ & $4.2\times10^{13}$  \\
Kepler-47d  & 0.085 & $9.5\times10^{6}$ & 0.905 & $1.4\times10^{-16}$ & $6.8\times10^{15}$ & 
$-1.4\times10^{-16}$ & $3.2\times10^{14}$  \\
Kepler-64   & 0.267 & $1.9\times10^{6}$ & 0.902 & $2.0\times10^{-15}$ & $3.2\times10^{14}$ & 
$-4.3\times10^{-15}$ & $1.2\times10^{13}$  \\
Kepler-413  & 0.286 & $2.4\times10^{6}$ & 0.767 & $3.6\times10^{-15}$ & $9.9\times10^{13}$ & 
$-6.0\times10^{-15}$ & $2.0\times10^{13}$  \\
Kepler-453  & 0.235 & $9.7\times10^{6}$ & 0.926 & $2.6\times10^{-16}$ & $3.4\times10^{15}$ & 
$-3.6\times10^{-16}$ & $1.1\times10^{14}$  \\
Kepler-1647 & 0.047 & $5.4\times10^{9}$ & 0.865 & $8.3\times10^{-19}$ & $3.3\times10^{18}$ & 
$-2.3\times10^{-19}$ & $2.5\times10^{17}$  \\
\hline
\end{tabular}}
\end{spacing}
\end{table*}

Table \ref{tab:kep} lists the physical parameters (mass and radius) of the binaries and the 
planets discovered by the {\it Kepler} mission, as well as the semimajor axes and eccentricities of 
the binaries (sub-index 1) and the planets (sub-index 2). The last column specifies the reference 
works from which the previous values were taken. In several cases, the difficulties in determining
the planetary mass has only allowed to establish an upper limit as defined by gravitational perturbation thresholds.

From Table \ref{tab:kep}, we can infer that most of the {\it{Kepler}} CBPs are in the  Neptune 
to Jupiter range, with expected tidal values $Q'_2=10^3-10^5$ \citep{Ferraz-Mello2013,Lainey2016}. 
For such planets, a CTL-model as the one adopted here should be a good approximation. 
On the other hand, there are some planets such as Kepler-47b and 
Kepler-47c that are probably closer to the super-earth range. In such systems, the direct application of our
model may not be adequate \citep{Efroimsky2012,Efroimsky2015}.

Although we are considering stars with very different masses and different possible internal 
structure, in the following sections we will assume the same value of the tidal parameters: 
$Q_0'=Q_1'=1 \times 10^6$.

\subsection{Tidal timescales}

In order to evaluate the importance of tidal evolution in these systems, we define characteristic 
timescales for the planetary spin evolution ($\tau_s$), the semimajor axis evolution ($\tau_a$) 
and eccentricity evolution ($\tau_e$) according to:
\be
\label{eq:taus}
\frac{1}{\tau_s} = \frac{1}{\Omega_2} \left<\frac{d\Omega_2}{dt}\right> \ \ \ ; \ \ \ 
\frac{1}{\tau_a} = \frac{1}{a_2} \left<\frac{da_2}{dt}\right> \ \ \ ; \ \ \ \frac{1}{\tau_e} = 
\frac{1}{e_2} \left<\frac{de_2}{dt}\right> .
\ee

In the case of the planetary spin, the simple form of Equation (\ref{eq:dodt}) allow us to 
accurately estimate the characteristic timescales of a CBP due to tidal effects as
\be
\tau_s = \frac{2 \mathcal{C}_2 a_2^6 }{3\mathcal{G}\mathcal{R}_2^5k_{2,2}\Delta t_2} \bigg(\sum_{i=0}^4 
K^{(s)}_i C^{(s)}_{i} \alpha^i\bigg)^{-1}. 
\label{eq26}
\ee

Results are presented in Table \ref{tab:kep_tid}, were the first numerical value is the current 
semimajor axis ratio $\alpha$ of the CBP. The two following columns give the characteristic 
timescale necessary for the planet to reach a stationary spin (i.e. $\tau_s$) and the 
corresponding equilibrium value $\left<\Omega_2\right>^{(\rm st)}$, expressed in units of its 
orbital frequency $n_2$. The planetary moment of inertia was taken equal to $\mathcal{C}_2 = 0.25 
m_2 \mathcal{R}_2^2$. 

The values of $\tau_s$ were calculated adopting $Q'_2 = 1$ so they should be scaled appropriately 
for other planetary tidal parameters. According to Equation (\ref{eq26}), $\tau_s \propto Q'_2$, 
making it straightforward to relate both quantities. In agreement with the numerical experiments 
done for Kepler-38 by \cite{Zoppetti2019}, pseudo-synchronisation appears to be attained rapidly in 
most {\it Kepler} systems even for large value for $Q'_2$. With the possible exception of 
Kepler-1647, we expect all {\it Kepler} CBP to currently lie in stationary spin-orbit 
configurations. As can be seen in the following column, all the stationary spins are expected to be 
sub-synchronous, some of them for a large amount (such as Kepler-35) and others almost in perfect 
synchronisation with its own mean motion (like Kepler-34).

The four last columns of Table \ref{tab:kep_tid} give the average time derivatives and 
characteristic timescales of the planetary semimajor axis and eccentricity. We assumed stationary 
values for $\Omega_2$ and current values for $a_2$ and $e_2$. Due to the complex functional 
dependence of the derivative with the initial conditions, the numerical values shown in the table 
should be considered local and not necessarily indicative of their primordial magnitudes. Even so, 
they give a qualitative idea of the strength of the tidal effects and, particularly, the sign of 
the present-day orbital variation.

\begin{figure*}[ht!]
\centering
\includegraphics[width=0.95\textwidth,clip]{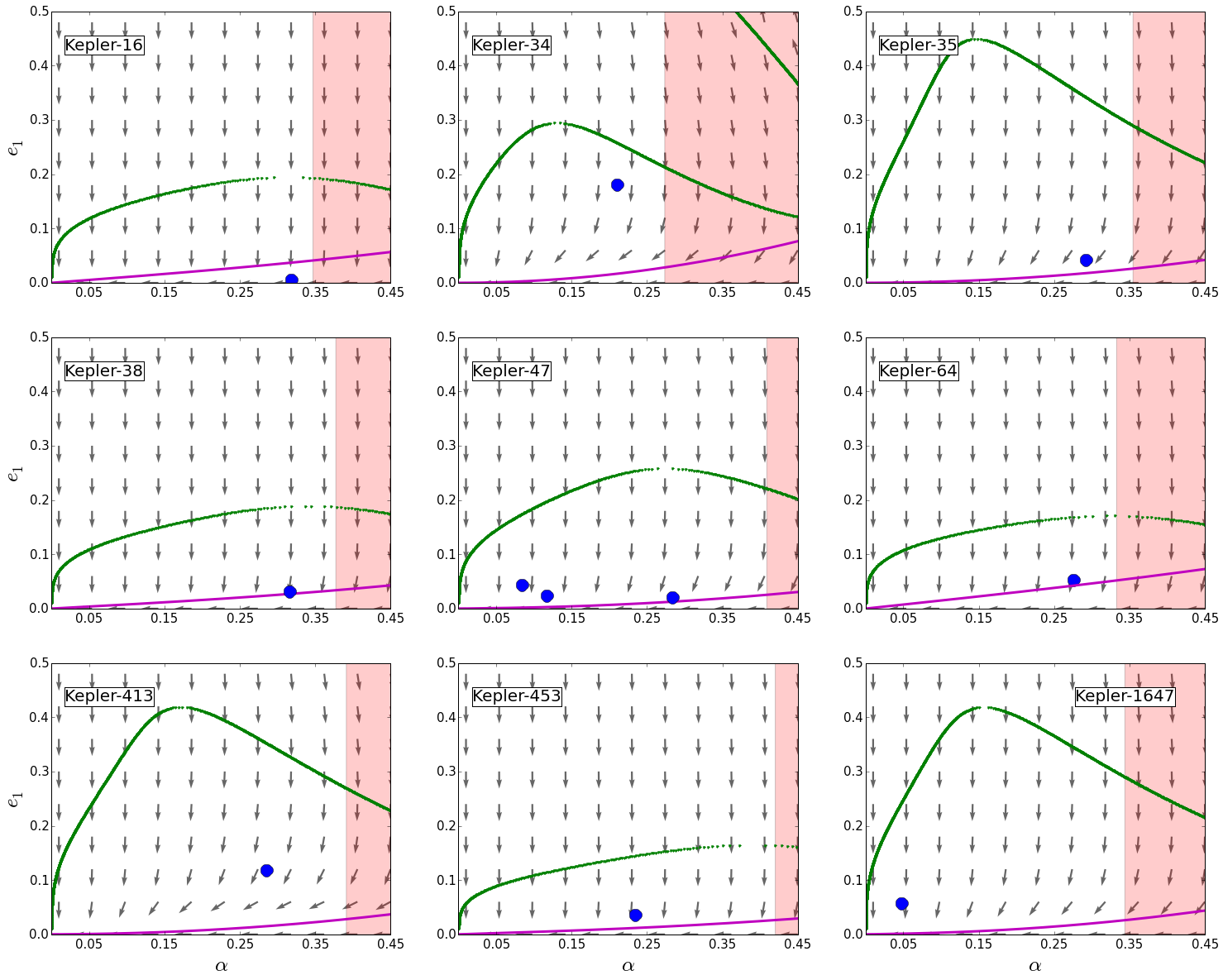}
\caption{Velocity vector fields in the $(\alpha,e_2)$ plane depicting the routes of tidal 
evolution in the vicinity of each {\it Kepler} CBP. The arrows show the direction of orbital 
evolution throughout the plane, although its size was kept constant and is thus not representative 
of the magnitude of the derivatives. The green curves corresponds to $\left< da_2/dt \right> = 0$) 
while the magenta curves indicate the location of the capture eccentricity $e_{\rm cap}$ as 
defined by equation (\ref{ec:ecap}). The pale pink region correspond to values of the orbital 
elements identified as unstable according to the criterion by \cite{Holman1999}. Current positions 
of the {\it Kepler} CBPs are shown in blue circles.}
\label{fig:kep}
\end{figure*}

For the observed orbital and physical parameters, our model predicts that all {\it Kepler} 
circumbinary planets should be migrating outwards as a consequence of their tidal interactions with 
the binary. However, the timescale necessary for a significant orbital migration is much larger 
than the age of the host star, even adopting very small values of $Q'_2$. Conversely, tidal 
evolution always seems to damp the planetary eccentricity, and the circularization process appears 
to occur in a slightly smaller timescale than the semimajor axis.

A final word on the case of Kepler-1647. Due to its large distance from the binary, as expressed by 
its small value of $\alpha$, the characteristic timescales for tidal evolution are typically three 
orders of magnitude larger than in any other system. Even so however, it is still possible for the 
planet to be close to a spin-orbit stationary solution even for moderate-to-large values of $Q'_2$.

\subsection{A geometrical interpretation for outward migration}
\label{sec:inte}

The prediction that all {\it Kepler} CBPs should be experiencing outward migration from the 
tidal effect raises the question as to why this happens. Although a full explanation is beyond 
the scope of this paper, we can present a simple geometrical interpretation that may give some 
insight. 

We consider the case in which the binary and the planetary orbit are circular, 
$e_1=e_2=0$. In addition, we assume that the planet is sufficiently far from the binary to neglect 
all terms in Equation (\ref{eq:dadt_ana}) explicitly dependent on $\alpha$. In this approximation, 
the stationary spins are trivial and equal to the mean motion: $\left<\Omega^{(\rm 
st)}_0\right>=\left<\Omega^{(\rm st)}_1\right> = n_1$ and $\left<\Omega^{(\rm st)}_2\right>= n_2$. 
From expression in Appendix \ref{ap:csem}, we find that the time derivative of the planetary 
semimajor axis acquires the simple form:
\be
\frac{1}{a_2} \left<\frac{da_2}{dt}\right> = \frac{6 n_2 \, m_2}{m_0+m_1} \bigg[ k_{2,0}\Delta t_0 
\bigg(\frac{\mathcal{R}_0}{a_2}\bigg)^5 + k_{2,1}\Delta t_1 
\bigg(\frac{\mathcal{R}_1}{a_2}\bigg)^5 
\bigg] (n_1-n_2) .
\label{eq:a2_cir}
\ee
We can compare this expression with that obtained in the two-body problem for a point-mass planets 
of mass $m_2$ in a circular orbit around a star $m_0$ rotating with arbitrary spin $\Omega_0$. In 
such a case, the semimajor axis of the planet evolves according to 
\be
\frac{1}{a_2}\frac{da_2}{dt} = \frac{6 n_2 m_2}{m_0} \bigg[ k_{2,0}\Delta t_0 \bigg( 
\frac{\mathcal{R}_0}{a_2}\bigg)^5 \bigg] \, (\Omega_0-n_2),
\label{eq:a2_cir_hut}
\ee
as can be seen, for example, in \cite{Hut1980}. The similarity between these two equations helps
explain why CBPs with low-to-moderate eccentricities migrate outwards. In the circumbinary 
geometry, the tidal effect of the binary star system may be substituted by a single body with a 
rotational frequency equal to the binary mean motion $n_1$. Since this quantity is always greater 
than the mean motion of the planet $n_2$, the semimajor axis of the planet increases. This 
situation is analogous to the two-body problem in which the central star $m_0$ rotates faster than 
orbital motion of the planet.

\subsection{Past tidal evolution}

Our next step is to attempt to reconstruct the past tidal evolution of the {\it Kepler} CB 
systems, as well as estimate their future trends. We assume that all bodies (stars and planet) are 
in stationary spin-orbit configurations and analyze only the changes in semimajor axis and 
eccentricity. Since the mass or radius of some of the planets are not well constrained, we 
completed the necessary physical parameters with the empirical mass-radius relation defined by
\cite{Mills2017}.

Figure \ref{fig:kep} shows, for each {\it Kepler} CBP, the tidal evolution velocity field around 
the current location of the planet in the $(\alpha,e_2)$ plane. The velocity field was computed 
assuming stationary spins for all intervening bodies and disregarding the tidal evolution of 
the binary stars. The arrows represent the direction of orbital evolution in $\alpha$ and $e_2$, 
and its size was kept constant.

The green curves show initial conditions with $\left< da_2/dt \right> = 0$. All points on the plane
above display inward migration ($da_2/dt <0$) while those below the curve lead to outward tidal 
migration. Kepler-34 shows a second green curve for large values of $\alpha$ that is also observed 
for some of the other systems for even higher values of $\alpha$ (which lie outside of the range 
adopted in the panels of Figure (\ref{fig:kep})). However, above such a curve, our model predicts
an increase in the eccentricity of the synchronised CBP. We suspect that this may be a 
spurious effect consequence of the truncation to fourth-order in the semimajor 
axis ratio when estimating the stationary CBP spin Equation (\ref{eq:omeq_ana}). 

The magenta curves show the capture eccentricity $e_{\rm 2,cap}$ as function of $\alpha$ for each 
system. Its value is given by the mean-square average between the forced eccentricity 
\citep{Moriwaki2004} and the mean eccentricity calculated for zero-amplitude secular variations 
\citep{Paardekooper2012}. Explicitly, 
\be
\label{ec:ecap}
e_{\rm 2,cap} = \sqrt{e_{2,f}^2 + e_{2,P}^2} \ ,
\ee
where the classical forced eccentricity is given by 
\be
e_{2,f} = \frac{5}{4}\alpha \frac{(m_0-m_1)}{(m_0+m_1)} 
e_1\frac{(1+\frac{3}{4}e_1^2)}{(1+\frac{3}{2}e_1^2)},
\ee
while the mean eccentricity takes the form
\be
e_{2,P} = \frac{3}{4}\alpha^2 \frac{m_0 m_1}{(m_0+m_1)^2} \sqrt{1 + \frac{34}{3}e_1^2}.
\ee
More details of the calculations leading to these values may be found in \cite{Zoppetti2019b}.
This so-called capture eccentricity defines the average value expected as the result of 
non-conservative exterior forces acting on the system.

Finally, initial conditions inside the pink region are dynamically unstable according to the 
empirical criterion of \cite{Holman1999}. Note however, that this criterion is only valid for 
small-mass planets and circular orbits, so their extent should be considered more qualitative than 
accurate.

Analysing the velocity vector fields, we first note that the eccentricity usually appears more 
affected than the semimajor axis, except for quasi-circular orbits where the opposite occurs. 
When the gravitational perturbations are included in the model, it is possible to observe that the 
orbital evolution of the system follows the locus of $e_{\rm 2,cap}$ as function of $\alpha$, as 
observed in \cite{Zoppetti2018} for Kepler-38.

In some systems we also observe that the domain associated to outward migration can reach high 
eccentricities (e.g. Kepler-35, Kepler-413, and Kepler-1647) while in others it is restricted to 
more circular trajectories (e.g. Kepler-16, Kepler-38, and Kepler-64). An explanation for such 
dichotomy may be found in \cite{Zoppetti2019}, where we found that in the low-eccentric systems, 
the size of the outward migrating-CB-region is directly proportional to the reduced mass 
$\bar{\mu}$ and inversely proportional to $e_2$. It is straightforward to check from Table 
\ref{tab:kep} that Kepler-35, Kepler-413 and Kepler-1647 satisfy quite these conditions while, for 
example, in the rest of the systems the value of $\bar{\mu}$ is significantly lower.

The proximity of the observed {\it Kepler} circumbinary planets to the stability limit with the 
binary has been the subject of several recent studies \citep[e.g.][]{Quarles2018}. While it has 
been proposed that this pile-up is unlikely to be affected by observational bias \citep{Li2016}, 
the recent discovery of a third and outer planet in Kepler-47 \citep{Orosz2019} may indicate that 
we are only detecting the tail of the distribution. Nevertheless, we still need to address the 
current population and its past orbital evolution. 

A key aspect of this question is the tidal evolution of the binary system and, particularly, what 
were the primordial orbital separation and eccentricity of the binary at the time of the planet's 
formation and migration. In \cite{Zoppetti2018} we showed that even for moderate values of the 
stellar tidal parameters the original binary eccentricity could have ben much larger, thereby 
pushing the instability barrier closer to the planet. Similar results were also found for 
Kepler-34. An outward tidal migration of the planet itself would have worked in the same direction 
leading to a potentially more unstable primordial configuration. 

These results could be interpreted as evidence that the orginal location of the CBPs was closely 
tied to the instability barried, perhaps in the form of sub-dense inner gaps in the protoplanetary 
disks. If so, then planetary traps such as those proposed by \cite{Kley2014} and \cite{Thun2018}
would be a more probable stalling mechanism for planetary migration than resonance capture.

Additionally, it is interesting to note that most of the CBPs located tightly packed to the 
binaries have eccentricities also very close to the capture eccentricity $e_{\rm 2,cap}$, with the 
exceptions of Kepler-34 and Kepler-413. Recently, \cite{Thun2018} suggested that low-mass CBPs are 
strongly influenced by the protoplanetary disc and their eccentricities may have been very excited 
during the migration process. If this hypothesis is confirmed, then another non-conservative effect 
must be invoked to explain their current state. Tidal interactions appears as a possibility, even 
if they would need to be extremely accentuated. Curiously, Kepler-35 and Kepler-38, two systems 
with $e_2 \sim e_{\rm 2,cap}$ belong to very old systems with estimated ages of the order of $\sim 10$ 
Gyrs \citep{Welsh2012, Orosz2012}. Perhaps accumulated tidal effects over such a long time could 
be at least partially responsible, and could help in constraining the magnitudes of tidal 
parameters for both stars and planets.

\section{Summary and discussion}
\label{sec:conclu}

We presented an extended version of the analytical tidal model for CB systems introduced in 
\cite{Zoppetti2019}. Once again, we assumed that all the bodies are extended and viscous in such a 
way that all interactions between pairs are considered but under the weak-friction regime, in which 
the tidal forces can be approximated by the classical expressions of \cite{Mignard1979}.

Starting from the variational equations of the spin and orbital evolution of the planet, we 
constructed an analytical approach by averaging over the mean longitudes. In all the cases, we 
expanded up to fourth-order in the semimajor axes ratio $\alpha$ but obtained closed expressions 
in the binary and planetary eccentricities. The resulting analytical model was compared with the 
results of numerical simulations and a very good agreement was found for all eccentricities and 
semimajor axes ratios up to $\alpha \sim 0.35$. 

Having expressions closed in eccentricities allowed us to apply the model to all well known
{\it Kepler} CBPs. We investigated their past tidal evolution in two steps. First, we calculated 
the characteristic tidal timescales and found that the typical time required for a CBP to acquire 
its stationary spin is typically much lower than the expected age of the host star, even for 
moderate-to-large values of $Q'$. Consequently, most of the observed system should lie in 
spin-orbital stationary solutions with a sub-synchronous rotational frequency. Regarding the 
orbital evolution we found that the typical tidal timescales in the {\it Kepler} systems are much 
longer than the rotational timescales, and little tidal induced orbital migration is expected to 
have occurred. The eccentricity damping timescales, however, are one or two orders of magnitude 
lower than $\tau_a$ and tidal interactions could have caused some decrease in the eccentricities 
during the systems' age.

In a second part of the work, we obtained some insight on the past tidal histories studying the 
tidal velocity fields around each planet. We found that all bodies are located in a tidal stream 
associated to an outward migration, pushing the planet away from the binary. Furthermore, with the 
exception of Kepler-34, all other systems are located distant from the curve separating inward 
and outward migrations. This seems to indicate that most or all of the systems lifetime was 
spent inside the region of outward migration. Consequently, their primordial separation from 
the binary should have been smaller. 

\cite{Thun2018} recently suggested that low-mass CBPs may have suffered strong excitation from the 
protoplanetary disk, leading to a highly eccentric final orbit. This contrasts with the current 
values which, in most cases, are found very close to the capture eccentricity. Perhaps tidal 
evolution played some role in damping the primordial value, which would help define some 
constraints on the numerical values of the tidal parameters of the system. 

We mention, finally, that most of the CBPs discovered so far by the {\it{Kepler}} mission
are in the range of Neptune to Jupiter masses, where CTL-models for tides should be valid.
However, we note that not much is known about the internal structure of this bodies and therefore
a precise estimation of the $Q'_2$-values becomes a difficult task.

\appendix
\section{Coefficients for the spin equations}
\label{ap:crot}

The variational equation for the planetary spin, up to fourth order in $\alpha$, was given in 
Equation (\ref{eq:dodt}), which is repeated here for convenience:
\be
\left<\frac{d\Omega_2}{dt}\right> = \frac{3\mathcal{G}\mathcal{R}_2^5k_{2,2}\Delta t_2}{2 
\mathcal{C}_2 a_2^6 } 
\sum_{i=0}^4 K^{(s)}_i \bigg(A^{(s)}_{i} n_1 + B^{(s)}_{i} n_2 + C^{(s)}_{i} \Omega_2\bigg) \, 
\alpha^i , \nonumber
\label{eq:dodt}
\ee
The $A^{(s)}_{i}$ coefficients, multiplying the mean motion $n_1$ of the binary, acquire the form
\bea
A^{(s)}_0 \hspace*{-0.25cm} &=& \hspace*{-0.25cm}  0  \nonumber \\
A^{(s)}_1 \hspace*{-0.25cm} &=& \hspace*{-0.25cm}  0  \nonumber \\
A^{(s)}_2 \hspace*{-0.25cm} &=& \hspace*{-0.25cm} -\frac{6 \sqrt{1-e_1^2}}{{(1-e_2^2)}^{13/2}} 
X_{1}  \\
A^{(s)}_3 \hspace*{-0.25cm} &=& \hspace*{-0.25cm} -\frac{168 \sqrt{1-e_1^2}}{{(1-e_2^2)}^{15/2}} 
X_{3} e_1 e_2 \cos(\Delta \varpi) \nonumber \\
A^{(s)}_4 \hspace*{-0.25cm} &=& \hspace*{-0.25cm} -\frac{6 \sqrt{1-e_1^2}}{{(1-e_2^2)}^{17/2}} 
\bigg( 8 Y_1 X_{5} + 175 X_{6} e_1^2 e_2^2 \cos(2 \Delta \varpi) \bigg) , \nonumber 
\eea
in terms of eccentricity functions $Y_i(e_1)$ and $X_i(e_2)$ which are explicitly given in 
Appendix \ref{ap:excf}. The coefficients that multiply the mean motion of the planet are found to 
be
\bea
B^{(s)}_0 \hspace*{-0.25cm} &=& \hspace*{-0.25cm}  \frac{2}{{(1-e_2^2)}^{6}} X_{1} \nonumber \\
B^{(s)}_1 \hspace*{-0.25cm} &=& \hspace*{-0.25cm}  \frac{72}{{(1-e_2^2)}^{7}} X_{3} e_1 e_2 
\cos(\Delta\varpi) \nonumber \\
B^{(s)}_2 \hspace*{-0.25cm} &=& \hspace*{-0.25cm}  \frac{3}{{(1-e_2^2)}^{8}} \bigg( 8 Y_1 X_{5}  + 
175 X_{6} e_1^2 e_2^2 \cos(2\Delta\varpi) \bigg)  \\
B^{(s)}_3 \hspace*{-0.25cm} &=& \hspace*{-0.25cm}  \frac{50 e_1 e_2}{{(1-e_2^2)}^{9}} \bigg( 24 Y_2 
X_{8} \cos(\Delta \varpi) + 49 X_{9} e_1^2 e_2^2 \cos(3\Delta\varpi) \bigg) \nonumber \\
B^{(s)}_4 \hspace*{-0.25cm} &=& \hspace*{-0.25cm} \frac{15}{{(1-e_2^2)}^{10}} \bigg( 8 Y_3 X_{11} + 
756 Y_4 X_{12} e_1^2 e_2^2 \cos(2 \Delta \varpi) \nonumber \\
          \hspace*{-0.25cm} &+& \hspace*{-0.25cm} \frac{9261}{16} X_{13} e_1^4 e_2^4 
\cos(4\Delta\varpi) \bigg) . \nonumber 
\eea
Finally, the coefficients accompanying the planetary spin rate are
\bea
C^{(s)}_0 \hspace*{-0.25cm} &=& \hspace*{-0.25cm} -\frac{2}{{(1-e_2^2)}^{9/2}} X_{2} \nonumber \\
C^{(s)}_1 \hspace*{-0.25cm} &=& \hspace*{-0.25cm} -\frac{45}{{(1-e_2^2)}^{11/2}} X_{4} e_1 e_2 
\cos(\Delta\varpi) \nonumber \\
C^{(s)}_2 \hspace*{-0.25cm} &=& \hspace*{-0.25cm} -\frac{9}{{(1-e_2^2)}^{13/2}} \bigg( 2 Y_1 X_{1} 
+ 25 X_{7} e_1^2 e_2^2 \cos(2\Delta\varpi) \bigg) \\
C^{(s)}_3 \hspace*{-0.25cm} &=& \hspace*{-0.25cm} -\frac{35e_1 e_2}{{(1-e_2^2)}^{15/2}} \bigg( 18 
Y_2 X_{3} \cos(\Delta \varpi) + \frac{175}{8} X_{10} e_1^2 e_2^2 \cos(3\Delta\varpi) \bigg) 
\nonumber \\
C^{(s)}_4 \hspace*{-0.25cm} &=& \hspace*{-0.25cm} -\frac{9}{{(1-e_2^2)}^{17/2}} \bigg( 8 Y_3 X_{5} 
+ 490 Y_4 X_{6}  e_1^2 e_2^2 \cos(2 \Delta \varpi) \nonumber \\ 
\hspace*{-0.25cm} &+& \hspace*{-0.25cm} \frac{3675}{16} X_{14} e_1^4 e_2^4 \cos(4\Delta\varpi) 
\bigg) . \nonumber 
\eea
We note that all coefficients are function of the eccentricities, while some are also dependent on 
the difference in the longitudes of pericenter. 

Recalling Equation (\ref{eq:omeq_ana}), the stationary spin rate predicted by our model may 
be written as
\begin{eqnarray}
\left< \Omega_2 \right>^{(\rm st)} = \sum_{i=0}^4 \left< \Omega_2 \right>^{(\rm st)}_i \alpha^i. 
\nonumber
\end{eqnarray}
The different order terms $\left< \Omega_2 \right>^{(\rm st)}_i$ are given by
\bea
\left< \Omega_2 \right>^{(\rm st)}_0 \hspace*{-0.25cm} &=& \hspace*{-0.25cm} -\frac{1}{C_0^{(s)}} 
D_0^{(s)} \nonumber  \\
\left< \Omega_2 \right>^{(\rm st)}_1 \hspace*{-0.25cm} &=& \hspace*{-0.25cm} 
-\frac{K_1^{(s)}}{K_0^{(s)}C_0^{(s)}} \bigg( D_1^{(s)} + C_1^{(s)} \left< \Omega_2 \right>^{(\rm 
st)}_0 \bigg) \nonumber \\
\left< \Omega_2 \right>^{(\rm st)}_2 \hspace*{-0.25cm} &=& \hspace*{-0.25cm} 
-\frac{K_2^{(s)}}{K_0^{(s)}C_0^{(s)}} \bigg( D_2^{(s)} + C_2^{(s)} \left< \Omega_2 \right>^{(\rm 
st)}_0 + \frac{K_1^{(s)}}{K_2^{(s)}} C_2^{(s)} \left< \Omega_2 \right>^{(\rm st)}_1 \bigg)  
\nonumber \\
\left< \Omega_2 \right>^{(\rm st)}_3 \hspace*{-0.25cm} &=& \hspace*{-0.25cm} 
-\frac{K_3^{(s)}}{K_0^{(s)}C_0^{(s)}} \bigg( D_3^{(s)} + C_3^{(s)} \left< \Omega_2 \right>^{(\rm 
st)}_0 + \frac{K_2^{(s)}}{K_3^{(s)}} C_2^{(s)} \left< \Omega_2 \right>^{(\rm st)}_1 \nonumber \\
 \hspace*{-0.25cm} &+& \hspace*{-0.25cm} \frac{K_1^{(s)}}{K_3^{(s)}} C_1^{(s)} \left< \Omega_2 
\right>^{(\rm st)}_2 \bigg) \\
\left< \Omega_2 \right>^{(\rm st)}_4 \hspace*{-0.25cm} &=& \hspace*{-0.25cm} 
-\frac{K_4^{(s)}}{K_0^{(s)}C_0^{(s)}} \bigg( D_4^{(s)} + C_4^{(s)} \left< \Omega_2 \right>^{(\rm 
st)}_0 + \frac{K_3^{(s)}}{K_4^{(s)}} C_3^{(s)} \left< \Omega_2 \right>^{(\rm st)}_1 \nonumber \\
\hspace*{-0.25cm} &+& \hspace*{-0.25cm}  \frac{K_2^{(s)}}{K_4^{(s)}} C_2^{(s)} \left< \Omega_2 
\right>^{(\rm st)}_2 + \frac{K_1^{(s)}}{K_4^{(s)}} C_1^{(s)} \left< \Omega_2 \right>^{(\rm 
st)}_3\bigg) \nonumber ,
\eea
where, for the sake of simplicity, we have defined new auxiliary functions 
$$D_i^{(s)} = A_i^{(s)} n_1 + B_i^{(s)} n_2.$$

\section{Coefficients for the tidal evolution of the semimajor axis and eccentricity}
\label{ap:csem}

As seen in Equation (\ref{eq:dadt_ana}), the time derivative of the planetary semimajor axis may 
be writen as
\be
\label{eq:dadt}
\left<\frac{da_2}{dt}\right> = \frac{ n_2}{\mathcal{G}m_2\sigma_1 a_2^4} \sum_{i=0}^4 
K_i^{(a)}\bigg(  A^{(a)}_{i} n_1 + B^{(a)}_{i} n_2  + C^{(a)}_{i} \bar\Omega_i^* \bigg) 
\, \alpha^i, 
\nonumber 
\ee
where the definitions of $K_i^{(a)}$ and $\bar\Omega_i^*$ were explicitly given in (\ref{eq:Ka}) 
and (\ref{eq:O*}). The coefficients are: 
\bea
A^{(a)}_0 \hspace*{-0.25cm} &=& \hspace*{-0.25cm} 0 \nonumber \\
A^{(a)}_1 \hspace*{-0.25cm} &=& \hspace*{-0.25cm} 0 \nonumber \\
A^{(a)}_2 \hspace*{-0.25cm} &=& \hspace*{-0.25cm} \frac{10 \sqrt{1-e_1^2}}{{(1-e_2^2)}^{8}} 
X_5  \nonumber \\
A^{(a)}_3 \hspace*{-0.25cm} &=& \hspace*{-0.25cm} \frac{400 \sqrt{1-e_1^2}}{{(1-e_2^2)}^{9}} 
X_{8}  e_1 e_2 \cos(\Delta \varpi) \\
A^{(a)}_4 \hspace*{-0.25cm} &=& \hspace*{-0.25cm} \frac{25 \sqrt{1-e_1^2}}{8 {(1-e_2^2)}^{10} 
} \bigg( 32 Y_1 X_{11} + 9 \bigg( 120 X_{12} -184 e_1^2 X_{21} + 127 e_1^4 X_{22} \bigg)\nonumber 
\\
\hspace*{-0.25cm} &\cdot& \hspace*{-0.25cm}  \frac{e_1^2 e_2^2 \cos(2 \Delta 
\varpi)}{{(1-e_1^2)}^{2}}  \bigg) \nonumber
\eea
\bea
B^{(a)}_0 \hspace*{-0.25cm} &=& \hspace*{-0.25cm} -\frac{2}{{(1-e_2^2)}^{15/2}} X_{15} \nonumber \\
B^{(a)}_1 \hspace*{-0.25cm} &=& \hspace*{-0.25cm} -\frac{114}{{(1-e_2^2)}^{17/2}} X_{16} e_1 
e_2 \cos(\Delta\varpi) \nonumber \\
B^{(a)}_2 \hspace*{-0.25cm} &=& \hspace*{-0.25cm} -\frac{1}{{(1-e_2^2)}^{19/2}} \bigg( 34 Y_1 
X_{17} + 1195 X_{18} e_1^2 e_2^2 \cos(2\Delta\varpi) \bigg) \nonumber \\
B^{(a)}_3 \hspace*{-0.25cm} &=& \hspace*{-0.25cm} -\frac{25 }{8 {(1-e_2^2)}^{21/2}} \bigg( 772 
Y_2 X_{19} e_1 e_2 \cos(\Delta \varpi) \\
\hspace*{-0.25cm} &+& \hspace*{-0.25cm} 2415 X_{20} e_1^3 e_2^3 \cos(3\Delta\varpi) 
\bigg) \nonumber \\
B^{(a)}_4 \hspace*{-0.25cm} &=& \hspace*{-0.25cm} -\frac{5 }{32 {(1-e_2^2)}^{23/2}} \bigg( 
1408 Y_3 X_{23} + 195300 Y_4 X_{24} e_1^2 e_2^2 \cos(2 \Delta \varpi) \nonumber \\
              \hspace*{-0.25cm} &+& \hspace*{-0.25cm} 222075 X_{25} e_1^4 e_2^4 
\cos(4\Delta\varpi) \bigg) \nonumber
\eea
\bea
C^{(a)}_0 \hspace*{-0.25cm} &=& \hspace*{-0.25cm} \frac{2}{{(1-e_2^2)}^{6}} X_{1} \nonumber \\
C^{(a)}_1 \hspace*{-0.25cm} &=& \hspace*{-0.25cm} \frac{72}{{(1-e_2^2)}^{7}} X_{3} e_1 e_2 
\cos(\Delta\varpi) \nonumber \\
C^{(a)}_2 \hspace*{-0.25cm} &=& \hspace*{-0.25cm} \frac{3}{{(1-e_2^2)}^{8}} \bigg( 8 Y_1 X_{5} + 
175 X_{7} e_1^2 e_2^2 \cos(2\Delta\varpi) \bigg) \\
C^{(a)}_3 \hspace*{-0.25cm} &=& \hspace*{-0.25cm} \frac{50 }{{(1-e_2^2)}^{9}} \bigg( 24 Y_2 X_{8} 
e_1 e_2 \cos(\Delta \varpi) + 49 X_{9} e_1^3 e_2^3 \cos(3\Delta\varpi) \bigg) \nonumber \\
C^{(a)}_4 \hspace*{-0.25cm} &=& \hspace*{-0.25cm} \frac{5}{16{(1-e_2^2)}^{10}} \bigg( 384 Y_3 
X_{11} + 36288 Y_4 X_{12}  e_1^2 e_2^2 \cos(2 \Delta \varpi) \nonumber \\
              \hspace*{-0.25cm} &+& \hspace*{-0.25cm} 27783 X_{13} e_1^4 e_2^4 \cos(4\Delta\varpi) 
\bigg) \nonumber .
\eea

In the same manner, the planetary eccentricity evolution up to fourth order in $\alpha$ is given 
by 
\be\label{eq:dedt}
\left<\frac{de_2^2}{dt}\right> = \frac{ n_2}{\mathcal{G}m_2\sigma_1 a_2^5} \sum_{i=0}^4 
K_i^{(e)}\bigg(  A^{(e)}_{i} n_1 + B^{(e)}_{i} n_2  + C^{(e)}_{i} \bar\Omega_i^* \bigg) \, 
\alpha^i 
\nonumber 
\ee
\label{eq19}
where $K_i^{(e)} = K_i^{(a)}$ and the coefficients are listed below:
\bea
A_0^{(e)} \hspace*{-0.25cm} &=& \hspace*{-0.25cm} 0 \nonumber \\
A_1^{(e)} \hspace*{-0.25cm} &=& \hspace*{-0.25cm} 0 \nonumber \\
A_2^{(e)} \hspace*{-0.25cm} &=& \hspace*{-0.25cm} \frac{75 \sqrt{1-e_1^2} } {(1-e_2^2)^7}X_{3} 
e_2^2 \nonumber \\
A_3^{(e)} \hspace*{-0.25cm} &=& \hspace*{-0.25cm} \frac{85 \sqrt{1-e_1^2} }{(1-e_2^2)^8} X_{30} e_1 
e_2 \cos(\Delta\varpi)  \\
A_4^{(e)} \hspace*{-0.25cm} &=& \hspace*{-0.25cm} \frac{25 \sqrt{1-e_1^2}} {8(1-e_2^2)^9 } \bigg[ 
304 Y_1 X_{8} e_2^2 \nonumber \\
            \hspace*{-0.25cm} &+& \hspace*{-0.25cm} \big( 380 X_{34} -256 e_1^2 X_{35} + 443 
e_1^4 X_{36} \big)  \frac{e_1^2 e_2^2 \cos(2 \Delta\varpi)}{(1-e_1^2)^{2}} \bigg] \nonumber
\eea
\bea
B_0^{(e)} \hspace*{-0.25cm} &=& \hspace*{-0.25cm} -\frac{18}{(1-e_2^2)^{13/2}}X_{3} e_2^2 
\nonumber \\
B_1^{(e)} \hspace*{-0.25cm} &=& \hspace*{-0.25cm} -\frac{27 }{ (1-e_2^2)^{15/2}}X_{26} e_1 e_2 
\cos(\Delta\varpi) \nonumber \\
B_2^{(e)} \hspace*{-0.25cm} &=& \hspace*{-0.25cm} -\frac{5 }{4(1-e_2^2)^{17/2}} \Big(296 Y_1 X_{8} 
e_2^2 + 361 X_{28} e_1^2 e_2^2  \cos(2 \Delta\varpi) \Bigg) \nonumber \\
B_3^{(e)} \hspace*{-0.25cm} &=& \hspace*{-0.25cm} -\frac{25 }{8(1-e_2^2)^{19/2}} \Big( 148 Y_2 
X_{31} e_1 e_2 \cos(\Delta\varpi) \\ 
            \hspace*{-0.25cm} &+& \hspace*{-0.25cm} 1141 X_{32} e_1^3 e_2^3 \cos(3 \Delta\varpi) 
\Big) \nonumber \\
B_4^{(e)} \hspace*{-0.25cm} &=& \hspace*{-0.25cm} -\frac{15 }{8(1-e_2^2)^{21/2}} \Big( 1504 Y_3 
X_{37} e_2^2 \nonumber \\ 
            \hspace*{-0.25cm} &+& \hspace*{-0.25cm} 5187 Y_4 X_{38} e_1^2 e_2^2 \cos(2 
\Delta\varpi) + 10017 X_{39} e_1^4 e_2^4 \cos(4 \Delta\varpi) \Big) \nonumber
\eea
\bea
C_0^{(e)} \hspace*{-0.25cm} &=& \hspace*{-0.25cm} \frac{11}{(1-e_2^2)^5}X_{4} e_2^2 \nonumber \\
C_1^{(e)} \hspace*{-0.25cm} &=& \hspace*{-0.25cm} \frac{39 } {2(1-e_2^2)^6}X_{27} e_1 e_2 
\cos(\Delta\varpi) \nonumber \\
C_2^{(e)} \hspace*{-0.25cm} &=& \hspace*{-0.25cm} \frac{45 }{(1-e_2^2)^7}  \Big( 4  Y_1 X_{3} e_2^2 
 + 5 X_{29} e_1^2 e_2^2  \cos(2 \Delta\varpi) \Bigg) \nonumber \\
C_3^{(e)} \hspace*{-0.25cm} &=& \hspace*{-0.25cm} \frac{85}{16 (1-e_2^2)^{8}} \Big( 48 Y_2 X_{30} 
e_1 e_2 \cos(\Delta\varpi) \\
            \hspace*{-0.25cm} &+& \hspace*{-0.25cm} 145 X_{33} e_1^3 e_2^3 \cos(3 \Delta\varpi) 
\Big) \nonumber \\
C_4^{(e)} \hspace*{-0.25cm} &=& \hspace*{-0.25cm} \frac{285}{8(1-e_2^2)^9} \Big( \frac{2224}{57} 
Y_3 X_{8} e_2^2 \nonumber \\
    \hspace*{-0.25cm} &+& \hspace*{-0.25cm} 112 Y_4 X_{34} e_1^2 e_2^2 \cos(2 \Delta\varpi) 
    + 147 X_{40} e_1^4 e_2^4 \cos(4 \Delta\varpi)  \Big) \nonumber .
\eea

\section{Eccentricity functions}
\label{ap:excf}

The eccentricity functions $Y_i(e_1)$ and $X_i(e_2)$ to which we refer in the previous equations 
are found to be combinations of the Hansen coefficients. Explictly, the first are found to be:
\bea
Y_1 = 1 + \frac{3}{2}e_1^2 \hspace*{1.3cm} &;& \hspace*{0.3cm} Y_2 = 1 + \frac{3}{4}e_1^2 \\
Y_3 = 1 + 5e_1^2+\frac{15}{8} e_1^4 \ \ \ &;& \hspace*{0.3cm} Y_4 = 1 + \frac{1}{2}e_1^2 \nonumber,
\eea
while the functions of the planetary eccentricity are given in Table \ref{tab:e2}:
\begin{table*}
\centering
\caption{Planetary eccentricity functions}
\label{tab:e2}
\begin{spacing}{1.5}
\begin{tabular}{l|l||l|l}
\hline \hline
$i$   &                $X_i$        &             $i$             &             $X_i$ \\
\hline
1      &   $1+\frac{15}{2}e_2^2+\frac{45}{8}e_2^4+\frac{5}{16}e_2^6$  & 21      &  
$1+\frac{91}{23}e_2^2+\frac{70}{23}e_2^4+\frac{175}{368}e_2^6+\frac{21}{2944}e_2^8$ \\
2      &   $1+3e_2^2+\frac{3}{8}e_2^4$ & 22      & 
$1+\frac{609}{127}e_2^2+\frac{2345}{508}e_2^4+\frac{1925}{2032}e_2^6+\frac{329}{16256}e_2^8$ \\
3      &   $1+\frac{15}{4}e_2^2+\frac{15}{8}e_2^4+\frac{5}{64}e_2^6$  & 23      & 
$1+\frac{755}{22}e_2^2+\frac{2205}{11}e_2^4+\frac{28455}{88}e_2^6+\frac{221025}{1408}e_2^8 
+\frac{54873}{2816}e_2^{10}+\frac{1575}{5632}e_2^{12}$ \\
4      &   $1+\frac{3}{2}e_2^2+\frac{1}{8}e_2^4$ & 24   & 
$1+\frac{242}{31}e_2^2+\frac{3521}{248}e_2^4+\frac{1827}{248}e_2^6+\frac{3787}{3968}e_2^8 
+\frac{7}{496}e_2^{10}$ \\
5      &   $1+14 e_2^2+\frac{105}{4}e_2^4+\frac{35}{4}e_2^6+\frac{35}{128}e_2^8$ & 25   & 
$1+\frac{1379}{470}e_2^2+\frac{721}{376}e_2^4+\frac{215}{752}e_2^6+\frac{7}{1504}e_2^8$ \\
6      &   $1+\frac{5}{2}e_2^2+\frac{15}{16}e_2^4+\frac{1}{32}e_2^6$  & 26   & $1+\frac{829}{36} 
e_2^2+\frac{3515}{72} e_2^4+\frac{3305}{192} e_2^6+\frac{5}{9} e_2^8$  \\
7      &   $1+e_2^2+\frac{1}{16}e_2^4$  & 27   & $1+\frac{25}{2} e_2^2+\frac{85}{8} 
e_2^4+\frac{5}{8} e_2^6$ \\
8      &   $1+7e_2^2+\frac{35}{4}e_2^4+\frac{35}{16}e_2^6+\frac{7}{128}e_2^8$ & 28   & 
$1+\frac{7707}{722} e_2^2+\frac{90335}{5776} e_2^4+\frac{49147}{11552} 
e_2^6+\frac{1295}{11552} e_2^8$ \\
9      &   $1+\frac{15}{8}e_2^2+\frac{9}{16}e_2^4+\frac{1}{64}e_2^6$  & 29   &  $1+\frac{35}{6} 
e_2^2+\frac{55}{16} e_2^4+\frac{5}{32} e_2^6$ \\
10      &  $1+\frac{3}{4}e_2^2+\frac{3}{80}e_2^4$  & 30   & $1+\frac{91}{4} e_2^2+\frac{385}{8} 
e_2^4+\frac{1085}{64} e_2^6+\frac{35}{64} e_2^8$ \\
11      &  $1 +\frac{45}{2} e_2^2 + \frac{315}{4} e_2^4 + \frac{525}{8} e_2^6 + \frac{1575}{128} 
e_2^8$ & 31   &  $1+\frac{1341}{37} e_2^2+\frac{21147}{148} e_2^4+\frac{74445}{592} 
e_2^6+\frac{115227}{4736} e_2^8$\\
12      &  $1 + \frac{14}{3} e_2^2 + \frac{35}{8} e_2^4 + \frac{7}{8} e_2^6 + \frac{7}{384} e_2^8$ 
& 32   & $1+\frac{8911}{1304} e_2^2+\frac{19887}{2608} e_2^4+\frac{17605}{10432} 
e_2^6+\frac{49}{1304} e_2^8$ \\
13      &  $1 + \frac{3}{2} e_2^2 + \frac{3}{8} e_2^4 + \frac{1}{112} e_2^6$ & 33   &  
$1+\frac{15}{4} e_2^2+\frac{27}{16} e_2^4+\frac{1}{16} e_2^6$ \\
14      &  $1 + \frac{3}{5} e_2^2 + \frac{1}{40} e_2^4$ & 34   & $1+\frac{21}{2} 
e_2^2+\frac{245}{16} e_2^4+\frac{133}{32} e_2^6+\frac{7}{64} e_2^8$ \\
15      &  $1+\frac{31}{2}e_2^2+\frac{255}{8}e_2^4+\frac{185}{16}e_2^6+\frac{25}{64}e_2^8$  & 35   
&  $1+\frac{1113}{64} e_2^2+\frac{14455}{512} e_2^4+\frac{8225}{1024} e_2^6+\frac{889}{4096} 
e_2^8$\\
16      &  $1+\frac{287}{38}e_2^2+\frac{385}{38}e_2^4+\frac{1645}{608}e_2^6+\frac{175}{2432}e_2^8$ 
& 36   & $1+\frac{4431}{443} e_2^2+\frac{6370}{443} e_2^4+\frac{27475}{7088} 
e_2^6+\frac{5761}{56704} e_2^8$ \\
17      &  $1+\frac{406}{17}e_2^2+\frac{6013}{68}e_2^4+\frac{5285}{68}e_2^6 + 
\frac{33355}{2176}e_2^8 +\frac{175}{544}e_2^{10}$  & 37   &  $1+\frac{45}{4} e_2^2+\frac{105}{4} 
e_2^4+\frac{525}{32} e_2^6+\frac{315}{128} e_2^8+\frac{21}{512} e_2^{10}$ \\
18      &  
$1+\frac{2373}{478}e_2^2+\frac{18865}{3824}e_2^4+\frac{7973}{7648}e_2^6+\frac{175}{7648}e_2^8$ & 38 
  &  $1+\frac{4114}{247} e_2^2+\frac{89257}{1976} e_2^4+\frac{60459}{1976} 
e_2^6+\frac{152579}{31616} e_2^8 +\frac{329}{3952} e_2^{10}$ \\
19      &  
$1+\frac{2277}{193}e_2^2+\frac{22239}{772}e_2^4+\frac{58065}{3088}e_2^6+\frac{72639}{24704}e_2^8 
+\frac{315}{6176}e_2^{10}$ & 39   & $1+\frac{1589}{318} e_2^2+\frac{5719}{1272} 
e_2^4+\frac{711}{848} e_2^6+\frac{329}{20352} e_2^8$ \\
20      &  $1+\frac{679}{184}e_2^2+\frac{1071}{368}e_2^4+\frac{749}{1472}e_2^6+\frac{7}{736}e_2^8$ 
& 40   & $1+\frac{11}{4} e_2^2+ e_2^4+\frac{1}{32} e_2^6$ \\
\hline
\end{tabular}
\end{spacing}
\end{table*}

\newpage
\begin{acknowledgements}
The authors are very grateful to M. Efroimsky for their helpful comments and suggestions.
This research was funded by CONICET, SECYT/UNC, and FONCYT.
\end{acknowledgements}

\bibliographystyle{aa}

\end{document}